\documentclass[sigconf,natbib=true]{acmart}
\usepackage{adjustbox}
\usepackage{tabularx}
\usepackage{float}
\usepackage{amsmath}
\usepackage{breqn}

\setcopyright{rightsretained}
\acmConference[SIGIR eCom'22]{ACM SIGIR Workshop on eCommerce}{July 15, 2022}{ Madrid, Spain}
\acmYear{2022}
\copyrightyear{2022}
\makeatletter
\renewcommand\@formatdoi[1]{\ignorespaces}
\makeatother
\acmISBN{}
\usepackage[labelfont=rm,textfont=rm]{subcaption}

\AtBeginDocument{%
  \providecommand\BibTeX{{%
    \normalfont B\kern-0.5em{\scshape i\kern-0.25em b}\kern-0.8em\TeX}}}

\usepackage[graphicx]{realboxes}
\usepackage{longtable}

\begin{document}

\title[Gender Stereotypes in Search for Children's Products]{Fire Dragon and Unicorn Princess}
\subtitle{Gender Stereotypes and Children's Products in Search Engine Responses}

\author{Amifa Raj}
\affiliation{%
  \institution{People and Information Research Team\\ Boise State University}
  \city{Boise}
  \state{Idaho}
  \country{USA}
  \postcode{83725-2055}
}
\email{amifaraj@u.boisestate.edu}

\author{Michael D. Ekstrand}
\affiliation{%
  \institution{People and Information Research Team\\ Boise State University}
  \city{Boise}
  \state{Idaho}
  \country{USA}
  \postcode{83725-2055}
}
\email{michaelekstrand@boisestate.edu}




\begin{abstract}
Search engines in e-commerce settings allow users to search, browse, and select items from a wide range of products available online including children's items. Children's products such as toys, books, and learning materials often have stereotype-based gender associations. 
Both academic research and public campaigns are working to promote stereotype-free childhood development.
However, to date, e-commerce search engines have not received as much attention as physical stores, product design, or marketing as a potential channel of gender stereotypes. To fill this gap, in this paper, we study the manifestations of gender stereotypes in e-commerce sites when responding to queries related to children’s products by exploring query suggestions and search results. We have three primary contributions. First, we provide an aggregated list of children's products with associated gender stereotypes from the existing body of research. Second, we provide preliminary methods for identifying and quantifying gender stereotypes in system's responses. Third, to show the importance of attending this problem, we identify the existence of gender stereotypes in query suggestions and search results across multiple e-commerce sites.
\end{abstract}


\keywords{information retrieval, gender stereotype, search results, query suggestions, children, e-commerce}

\maketitle

\section{Introduction}
Information retrieval (IR) systems such as search engines for e-commerce settings serve a wide range of users satisfying their various information needs for different context. Further, beyond their ability to meet a user's information need, there are number of potential social concerns raised by these systems, and a growing body of research is actively studying various aspects of the ethical \cite{milano2020recommender, urman2021earth}, bias \cite{otterbacher2018addressing}, and fairness \cite{ekstrand2021fairness} challenges in IR systems.
However, the impact of these ethical issues on children has been overlooked in the corresponding research. We are concerned with an important issue that has not yet seen much study: the impact of representational harms \cite{crawford17} in search engine responses, particularly the way gender stereotypes may be reflected to children.
\citet{noble2018algorithms} highlighted how search engines can reinforce negative gender, racial, and intersectional stereotypes when used by children, and the risk such exposure can cause; in this paper, we attempt to measure the presence and extent of gender stereotypes for children's products in search systems of e-commerce settings. 

\citet{ellemers2018gender} describes gender stereotypes as shared (often implicit) beliefs and social expectations associated with people based on their gender. One way this manifests for children is by categorizing toys based on the gender of the child expected to play with them, reinforcing social expectations and encouraging people to believe that products can (and should) be differentiated based on gender \cite{liben2018cognitive}. 
These unnecessary gender associations appear in marketing, packaging, and advertisement of children's products through a variety of means, including color, labeling, and images \cite{berenbaum2018beyond, fulcher2018building}. Public-interest campaigns have documented the presence of gender stereotypes in the promotion of children's products and the negative implications of this practice \cite{etheridge2015let, fawcett19}; psychology research has also studied the existence and impact of gender stereotypes in children's products \cite{kneeskern2020examining, seitz2020effects, skovcajic2020boys} showing the importance of considering this problem.

Most of the efforts on this to date have focused on marketing and store design, but these are not the only channels through which stereotypes may be expressed. Since various e-commerce systems are frequently used to advertise, sell, and market children's products, the product marketing and store design itself are not the only channels through which stereotypes may be expressed: the search engines in those e-commerce sites may also replicate and reinforce gender stereotypes associated with children's products. 

It is important to investigate whether and how these systems may perpetuate or reinforce gender stereotypes to understand the potential impact of the existence of gender stereotypes in this source and thus developing strategies to measure and mitigate the gender stereotypes associated with children's products. 
\citet{baker2013white} and \citet{noble2018algorithms} showed that search engines often stereotypically associate gender in their responses.
We previously argued the need to consider the role of IR in replicating and reinforcing gender stereotypes regarding children's products through additional examples specifically focusing on learning environments \citep{raj2021pink}. Building on that agenda, we provide initial quantitative results on detecting and exploring the presence of gender stereotypes associated with children's products in e-commerce search systems to further highlight the need to investigate the role of IR in reinforcing gender stereotypes, particularly for search tasks that affect children (either because children are directly using the system, or because adults are using it to find information and products for children).

Our goal is to identify how gender stereotypes may appear in \emph{query suggestions} and \emph{search results} when searching for children’s products in e-commerce settings. We examined multiple e-commerce systems and analyzed their responses for associations of gender with children's products, based on a collection of commonly gender-stereotyped children's products from previous research.









We have three main contributions:
\begin{itemize}
    \item We provide an aggregated list of children's products with associated gender stereotypes from existing research.
    \item  We provide a simple and explainable preliminary methods to quantify gender-product associations while exploring the existence and persistence of gender stereotypes associated with children's products in query suggestions and search results in e-commerce settings.
    \item We identify the potential existence of gender stereotypes in query suggestions and search results of search engines, across multiple e-commerce systems, showing the importance of attending to this problem.
\end{itemize}

This work is not primarily an audit study to make normative claims about any particular e-commerce systems, although we hope developers of such systems will attend to our results. We focus on providing empirical support for the existence of the problem across the e-commerce search landscape in general as a prompt for discussion and further research.

\section{Background}

Our work draws primarily from three streams: gender stereotypes in children's products, gender stereotypes in IR, and IR for children.

\subsection{Gender Stereotype in Children's Products}

The term ``stereotype'' refers to the concept of categorizing people into groups and associating popular beliefs or common patterns with those groups \citep{cauteruccio2020investigating}. Gender stereotypes are common beliefs and social expectations associated with particular genders \cite{ellemers2018gender, cauteruccio2020investigating}. 

Gender stereotypes have long been associated with children's products; these associations persist in society and are reinforced by the marketing, packaging, and advertising of children's products \cite{auster2012gender}. 
The trend of assuming gender appropriateness for toys leads to gender stereotypes associated with them. 
\citet{cherney2018characteristics} and \citet{blakemore2005characteristics} showed that characteristics such as attractive, shiny, domestic, nurturing, fashion-oriented, pretend-play, and pink or pastel colors are commonly associated with ``toys for girls'', while active, adventurous, violent, aggressive, spatial, movement-based, STEM, competitive, somewhat dangerous, and black or dark colored are associated with ''toys for boys''. Retailers and media reflect this categorization and further reinforce them through marketing \cite{auster2012gender}, advertising \cite{azmi2021gender}, and packaging \cite{souza2021boys}. Children are exposed to this concept of gender-based toys through their social interactions, including parents, peers, and signals in society at large \cite{freeman2007preschoolers, kollmayer2018parents, eisen2021parents}. 

\citet{weisgram2018gender} explored various aspects of gender stereotypes associated with children's products, analyzing the characteristics of gendered products and the causes and consequences of gender-typed playing. Early childhood exposure to gender stereotypes through playing or engaging with gender-stereotyped products has long term impact on physical, social, and cognitive development including future ability, skills, personality, social interaction, perception towards gender roles, and career interests \cite{fulcher2018working, dinella2018gender, fulcher2018building, spinner2018peer, weisgram2014pink}.

In addition to academic research on stereotypes and their effects, consumer campaigns have worked to break gender-stereotyped toy practices. Efforts such as \textit{Let Toys be Toys}\footnote{https://www.lettoysbetoys.org.uk/} and \textit{Smash Stereotypes}\footnote{https://www.fawcettsociety.org.uk/smashstereotypes} are working to challenge gender stereotypes in children’s products with the goal of providing children with a stereotype-free childhood. These campaigns have had impact; the popular retailer Target \cite{dinella2018gender} and manufacturer LEGO \cite{russell_2021} have both decided to eliminate gender-based categorization of their products. California has also enacted a law requiring gender-neutral sections in stores to reduce gender-based discrimination \cite{beam_2021}.


\subsection{Gender Stereotypes in IR}

Ethical and social concerns, such as fairness, receive significant research interest in IR \cite{ekstrand2019fairness, gao2021addressing, urman2021earth}. Several studies have specifically looked to identify, assess, and mitigate gender-based social biases in IR \cite{raj2020comparing, makhortykh2021detecting, epps2020artist, ekstrand2021exploring}. Much of this work, however, focuses on distributional harms: ensuring that providers and users of different genders receive equitable benefits from the system. Representational harms --- harms arising from how the system represents users or items, either internally or in its results ---  of IR is still under-studied; replicating and reinforcing gender stereotypes is one such harm.

In AI and machine learning more broadly, \citet{abbasi2019fairness} studied stereotypes as representation harm in machine learning pipelines and \citet{ahn2022effect} showed the effect of gender stereotypes in AI-recommended products by investigating human interaction with gendered AI agents. In IR, \citet{kay2015unequal} showed the existence of gender stereotypes in image search results for occupations, as evaluated by human participants, and \citet{magno2016stereotypes} found that image results may impose certain attractiveness stereotypes depending on the language used to formulate queries. This work has not considered the presence of gender stereotypes in textual representations, or in product results of search engines in e-commerce settings. The closest work to our own is that of \citet{fabris2020gender}, who looked at gender stereotypes in retrieved documents of search results and proposed a \textit{Gender Stereotype Reinforcement} measure leveraging Word2Vec's \cite{mikolov2013linguistic} tendency to reflect gender stereotypes in word embeddings. \citeauthor{fabris2020gender} evaluated 
well-known IR algorithms on the Robust04 TREC dataset \cite{voorhees2004overview}, detecting stereotypes in a controlled environment instead of production systems. 

While there is some progress on recognizing the presence of stereotypes in IR systems \cite{noble2018algorithms, baker2013white,roy2020age, sweeney2013discrimination}, there hasn't been enough work that can help to systematically detect, assess, and mitigate gender stereotypes in IR. The work to date has also focused on the system's primary results, such as search results in response to a query; how various phases across an IR workflow (such as search results and query suggestions) replicate and exhibit gender stereotypes is still unexplored, as is the the role of IR in replicating and reinforcing gender stereotypes associated with specific problem settings like searching for children's products has not received attention to our knowledge.
\subsection{IR for Children}
Most research on IR, including research on its ethical and social dimensions, usually assume a ``default'' user or society member, often implicitly assumed to be an adult. There is a line of work on building better web environments for children, particularly in search; this research often focuses on explicit content concerns such as inappropriate language or adult content \cite{madrazo2018looking, alghowinem2018safer}, or on better meeting children's information needs \cite{dragovic2016sven}.
Connecting the work on ethical and social aspects of IR, such as fairness \cite{ekstrand2021fairness}, with the particular needs and experiences of children is a significant gap in the literature. \citet{noble2018algorithms} showed how search systems can perpetuate negative stereotypes in a way that impacts children, but most work building on her argument has focused on the reproduction of harmful societal ideas, not on their particular impacts on children. Thus, the role of search systems in manifesting harmful expectations and beliefs to children through their response is still overlooked; filling this gap will help achieve the goal of building safer search environments for children.

In this work, we draw our attention to two important issues that have not yet seen much study: considering the impact (and resulting ethical considerations) of search systems specifically on children; and studying representational harms in their response, particularly the way gender stereotypes may be reflected. 




\section{Data and Methods}

To identify the existence and persistence of gender stereotypes associated with children's toys in search systems in e-commerce settings, we explore query suggestions and search results quantifying the gender associations that manifest in each of these stages. 
We seek simple and explainable methods to provide a first approach in quantifying gender-product associations to demonstrate their existence and serve as a foundation for future work that will refine and further develop relevant measurements.

To conduct our experiments, we explored two popular US e-commerce sites: \textit{Amazon}\footnote{https://www.amazon.com/}, because they are the most widely-used e-commerce site in the United States \cite{chevalier21} and \textit{Target}\footnote{https://www.target.com/}, because they have taken action to end gendered categorization of toys \cite{murnen2018fashion, dinella2018gender}, and we want to see if this is reflected in search engine responses.

\subsection{Documented Toy Gender Stereotypes}

To identify the presence of gender stereotypes associated with children's products in a systematic way that connects to existing research and advocacy, we needed to identify items that are often have stereotyped gender associations.  
Existing research and advocacy campaigns provide lists of commonly-held toy-gender associations, categorizing toys as being targeted towards boys or girls; we used four of these lists to build our stereotyped toy sets: \citet{richardson1982children}, \citet{blakemore2005characteristics}, \citet{murnen2016boys}, and
Smash Stereotypes by Fawcett \cite{fawcett19}.
We combined the toys listed for each gender category by each of these studies and reports and removed duplicates to obtain lists of boy-associated, girl-associated, and neutral toys; Table~\ref{tab:items} shows the resulting full toy lists.

\begin{table*}[]
\footnotesize
\centering
\begin{minipage}{\textwidth}
\caption{List of toys and gender associations from previous papers and campaigns.}
\label{tab:items}
\begin{tabular}{lp{6.4in}}
\toprule
Boys & 
toy vehicles; military toys; race cars; outer space toys; construction toys; car toys; video games; building blocks; dinosaur toys; lego cars; lego toys; depots; machines; doll-humanoid; action figures; gi joe action figure; spider-man; guitar; castle tent; microscope; weather forecasting toy; bug collection set; barn; grill; toy guns; volcano creator; pokemon cards; space station toys; basketball hoop; gears; train set; lincoln logs; police station toy; airport toy; police officer gear;  dragon ball z; superhero costume; helicopter; racetrack; remote controlled cars; football; hockey stick; sword toy; miniature guns; truck set; sully costume; woody valentine; spider man costume; batman costume; ninja costume; wwe action figure; construction toys; war toys;   educational material; computer games; spaceship lego; firefighter gear; soldier toys; space station; science project kit; toy rocket; soccer ball; toy robots; toy drones; monster trucks; nerf guns
\\ 
\midrule
Girls  & 
doll; domestic toys; educational art; clothes; dollhouses; clothing accessories; doll accessories; furnishing; ballerina costume; barbie costume; barbie doll; barbie jeep; play makeup; bratz doll; jewelry; my little pony; tea set; vanity set; princess costume; baby doll; beads; easy bake oven; baby doll stroller; sewing machine; pink ice skates; ken doll; toy kitchen; broom; beanie baby bear; toy vacuum cleaner; veterinarian kit; kitchen set; stuffed animals; plush toys; unicorn; sparkly toys; disney doll; baking kit; painting kit; dora doll; power ranger valentine; dorothy costume; minnie mouse; monster high doll; monster high costume; fashion doll; tiara; craft toys; beauty products; tea party set; pink doll; cupcake maker; makeup kit; fashion dolls; crochet kit; playhouse; pink ice skates; princess sword; doctors kit; skin care kit; barbie furniture set
\\
\midrule
Neutral        & 
toy animals; educational teaching; musical games; games; books; live animals; candy land; winnie the pooh; karaoke machine; elmo; gardening tools; crayons; doctor kit; tricycle; play-doh; leappad; trampoline; swing set; mr potato head; spongebob square pants; math flash card; wagon; bus; tree house; wood blocks; harry potter books; scooter; drum set; puzzles; board games; rock painting
\\
\bottomrule
\end{tabular}
\end{minipage}
\end{table*}



    
    
\begin{figure*}[!ht]
     \begin{subfigure}[b]{0.35\columnwidth}
     \includegraphics[width=0.9\textwidth]{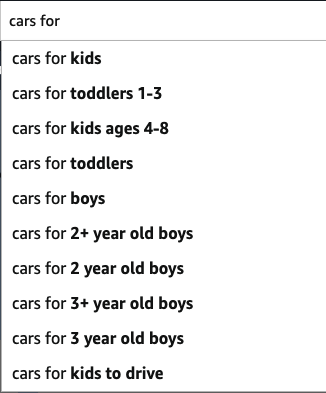}
     \caption{Query suggestion for ``cars for'' in Amazon}
     \label{fig:qs}
     \end{subfigure}
    \hspace{1em} 
    \begin{subfigure}[b]{0.6\columnwidth}
       \includegraphics[width=\textwidth]{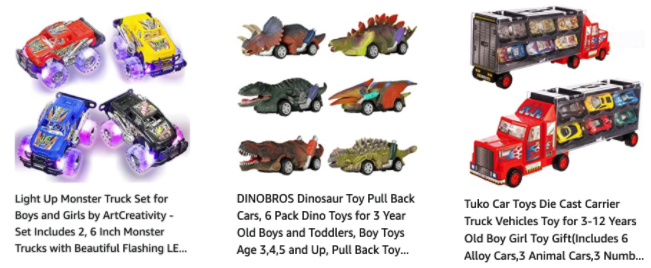}
       \caption{Search results for ``cars for boys'' from Amazon}
       \label{fig:car_boys}
   \end{subfigure}
   \hspace{1em}
   \begin{subfigure}[b]{0.6\columnwidth}
        \includegraphics[width=\textwidth]{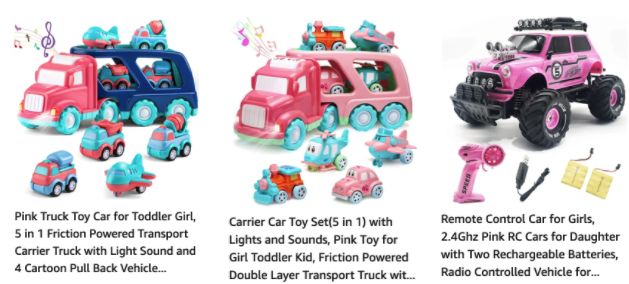}
        \caption{Search results for ``cars for girls'' from Amazon}
        \label{fig:car_girls}
    \end{subfigure}
    \caption{Query suggestions and search results for cars}
    \label{fig:amazon}
\end{figure*}

\subsection{Query suggestions}
\label{sec:exp-qs}

Many search engines provide \emph{query suggestions} to aid writing query process.  
The suggested query completions sometimes include gender markers; for example, figure~\ref{fig:qs} shows an e-commerce site suggesting ``cars for boys'' as a completion of ``cars for''. This is one way the e-commerce search systems may replicate existing gender stereotypes: the presence of the gendered term reflects an association between the initial query and gender, typically learned from underlying data. It may further reinforce those gender stereotypes by nudging users to select the gendered query.

We presented each item in Table~\ref{tab:items} as a query to \textit{Amazon} and \textit{Target} search systems and recorded the completions it suggested using the \texttt{fake-useragent}\footnote{https://pypi.org/project/fake-useragent/} package.
We then counted how often each system suggested a completion containing a gender to measure the gender association that system inferred for each product.
This setup allows us to identify if search systems associate target gender with items through query suggestions and whether those associations replicate previously-documented gender stereotypes.

\subsection{Search Results}
\label{sec:exp-sr}

When a search system retrieves results, it typically presents them in a ranked list sorted by their estimated relevance to the query \cite{schutze2008introduction}. One way the system may reflect gender associations and stereotypes is in its response to gendered queries: when a query explicitly mentions a gender, the system may retrieve a different set of relevant results than it would with a different gender, reflecting particular gender associations back to its users.
In addition to the products themselves, gender stereotypes can manifest in search results through various characteristics such as colors, keywords, terms, and features associated with the retrieved products. Figure~\ref{fig:car_boys} and figure~\ref{fig:car_girls} show top results for the queries ``cars for boys'' and ``cars for girls'', respectively showing difference in colors, products, features, and keywords with the change of gender in the query. Thus search results may reinforce and propagate gender stereotypes by reflecting unnecessary gender association with children's products.

We collected search results from \textit{Amazon} and \textit{Target} to explore their role in replicating and propagating gender stereotypes in children's products through the search results themselves. For each item in Table~\ref{tab:items}, we used \textit{Selenium}\footnote{https://www.selenium.dev/} to issue both gender-specific and gender-neutral queries, render the search result pages, and collect the titles of the retrieved products. 


To explore the differences in response to gender, we observe the difference in keywords associated with retrieved products when we issue a query with explicitly mentioning gender (gender-specific queries), either one of the items from Table~\ref{tab:items} (e.g. ``truck set for girls'') or a generic query (e.g. ``toys for boys''). For each query, we extracted frequent keywords from search results and identified unique terms that are commonly associated with products retrieved given a specific gender.
This strategy let us explore how systems replicate social stereotypes while retrieving gender-specific results.

To identify specific toy-gender associations in search results, we investigated the similarity between gender-neutral and gender-specific (gender mentioned in query) retrieved result sets for each item; if the neutral search results closely match with a specific gendered search results, we identify that product to be associated with that gender. For example, if the search results of the query ``dollhouse'' have more similarity with search results of ``dollhouse for girls'' than ``dollhouse for boys'', we assume the item (dollhouse) is to be associated with the female gender.


\section{Empirical Results}

We now turn to the results of this exploratory analysis and to that end, we answered two primary research questions:
\begin{description}
\item[RQ1.] Do query suggestions reflect gender stereotypes associated with children’s products?
\item[RQ2.] Do search results manifest gender stereotypes associated with children’s products?
\end{description}



\subsection{Query Suggestions}

Applying the collection strategy described in Sec.~\ref{sec:exp-qs} to the two e-commerce systems we studied yielded a total of 334 query suggestion sets. For a given system and a query, we collected the set of suggested queries. We then computed a girl-association score by counting the number of times ``girl'' or ``girls'' appears in the query suggestions (e.g. by expanding ``truck'' to ``truck for girls''); likewise we compute boy-association score by looking for ``boy'' and ``boys''. We used the difference between these two scores in determining the system associated gender label for each item.
Our investigation into this first question is in the form of three sub-questions.

\textit{\textbf{RQ1.a.} Do e-commerce search systems associate gender with children's products in query suggestions?}

\begin{figure}[!ht]
    \begin{subfigure}[t]{0.47\columnwidth}
            \includegraphics[width=\textwidth]{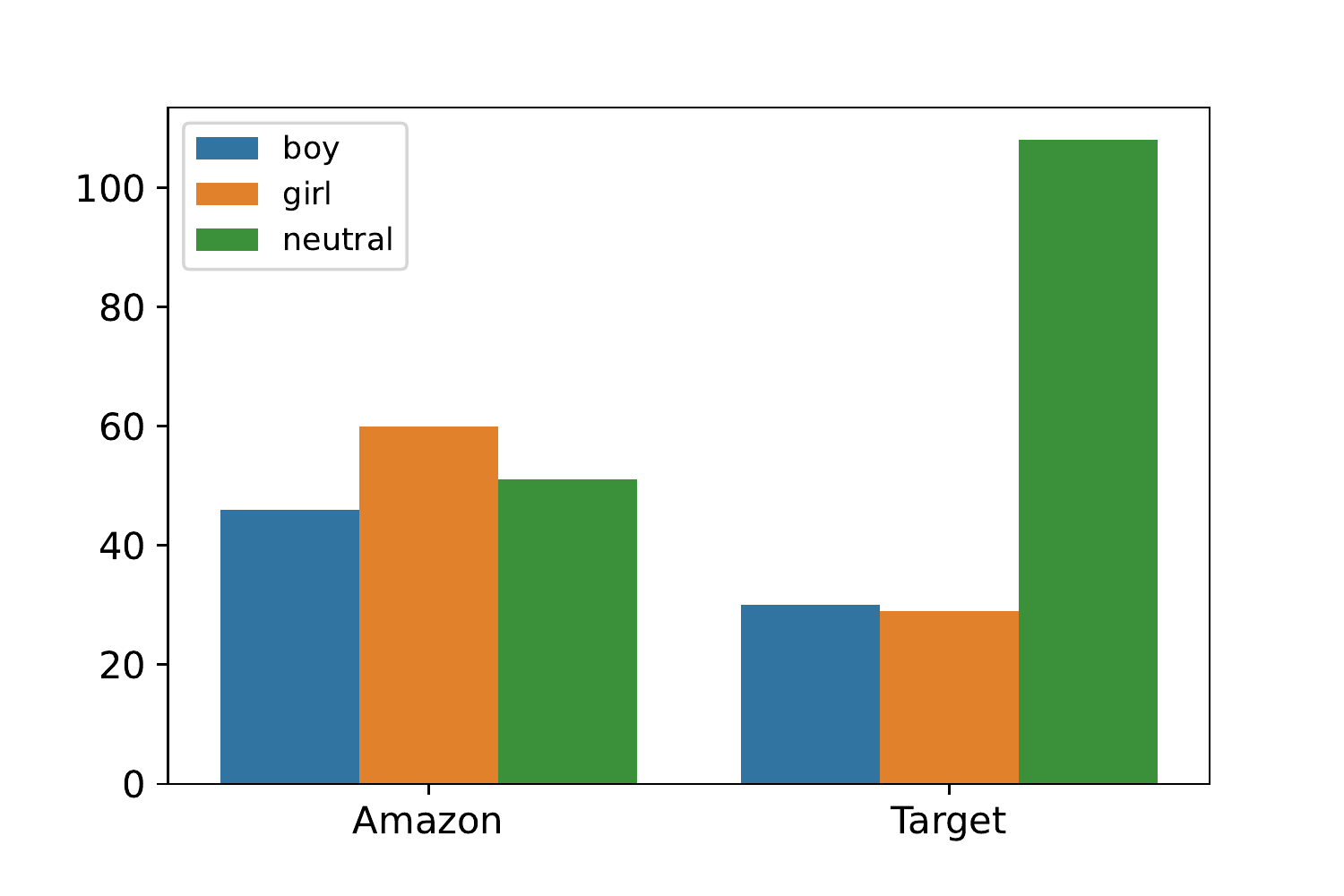}
    \caption{Item gender associations.}
    \label{fig:platform_indiv}
    \end{subfigure}
    \begin{subfigure}[t]{0.47\columnwidth}
        \includegraphics[width=\textwidth]{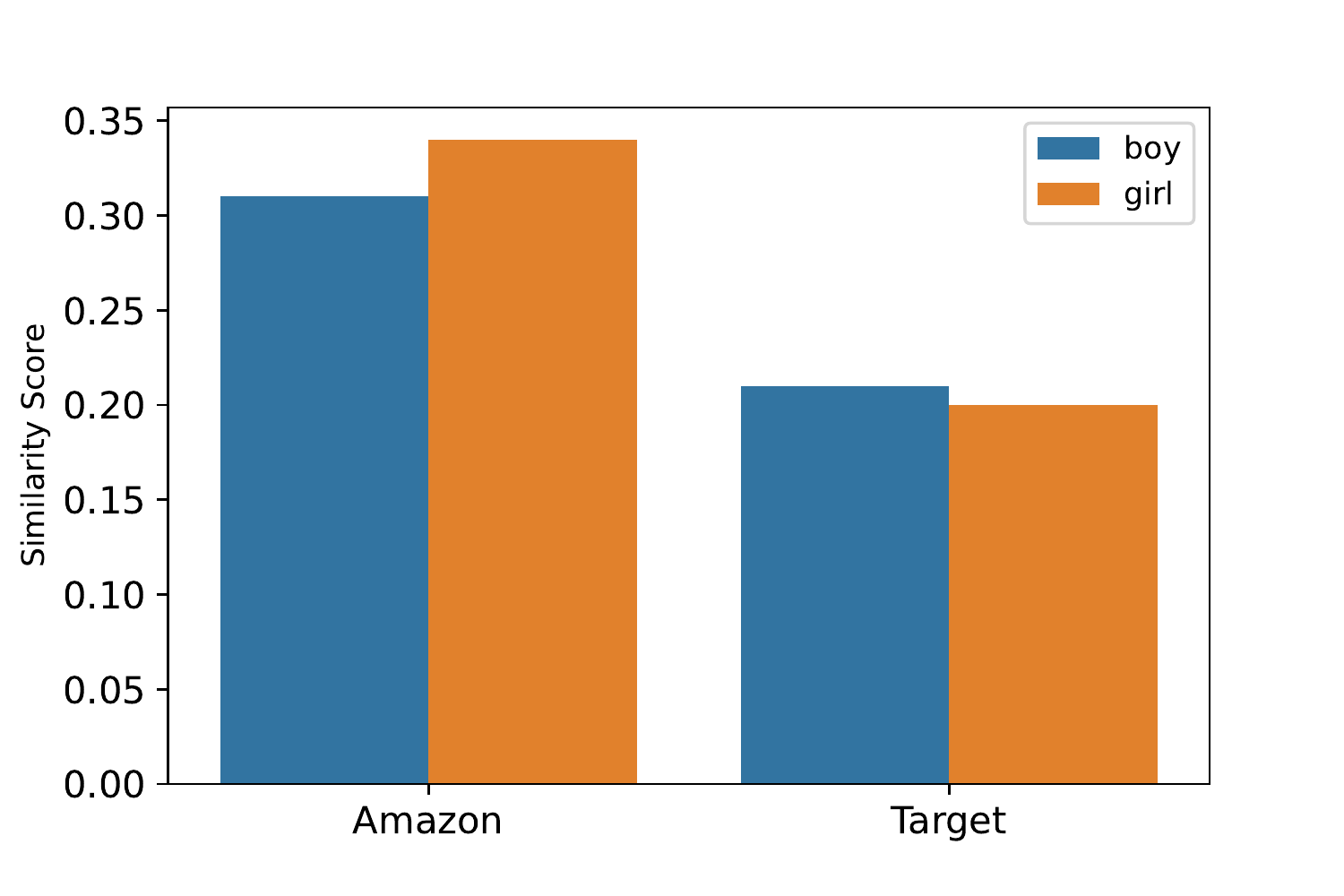}
    \caption{Similarity between system derived list and stereotyped toy list.}
    \label{fig:gender_match}
    \end{subfigure}
    
    \caption{Persistence of previously-documented gender stereotype associated with children's products across e-commerce query suggestions}
    \label{fig:platforms}
\end{figure}

To observe either system's tendency of attaching gender with children's products, we computed the number of items that were associated with genders by the systems while suggesting queries. Figure~\ref{fig:platform_indiv} shows that number of items each system associated with each gender. We first look at how likely the system is to associate a gender at all; Amazon had the highest tendency of attaching a gender to items through its query suggestions.

\textit{\textbf{RQ1.b.} Do e-commerce search systems replicate stereotypes associated with children's products through query suggestions?}

After identifying the tendency of associating gender with items through query suggestion, we wanted to explore how likely these systems reflect previously-documented labels for the items.
With that goal, we examined the individual lists of items with gender association derived from each system.

To quantify how each system \emph{replicates} gender stereotypes through its query suggestions, we compared the gender associations in its query associations with the previously-documented stereotypes. For each system, we calculated the \textit{Jaccard similarity} between sets of items the system associates with a gender and the set of items for that gender from the prior research (Table~\ref{tab:items}).
Figure~\ref{fig:gender_match} shows the similarity score for each system.
Both system's propagation of stereotype was statistically significantly greater than expected by chance ($p<10^{-5}$ with a one-tailed randomization test\footnote{Our sets of items are not representative or random samples of the space of possible items, so most statistical tests are not appropriate; the randomization test's null hypothesis that the labels are independent of other item attributes is more appropriate for this analysis. For each system, we shuffled the system-generated gender label of items and calculated the similarity score between the item list for a gender and the pre-documented item list for that gender, thus simulating the distribution of similarities under the null hypothesis.}). This provides evidence that these systems are replicating the gender stereotype associations of our toy list in their query suggestions.

Our results show that query suggestions appear to reproduce previously-documented gender stereotypes to some extent, in that the stereotyped gender is more likely to appear in query suggestions when searching for that item.

\textit{\textbf{RQ1.c} Does the prevalence of gender stereotypes vary between query suggestions of different systems?}

After observing the overall tendency of e-commerce search system to replicate gender stereotypes through query suggestions, we now turn to whether the systems differ in their responses. 

From figure~\ref{fig:gender_match}, we observe that Amazon and Target have similar query suggestion rates (fraction of items that have suggestions) but differ in their tendency of associating gender with items at all, in addition to differing in their replication of previous gender stereotypes through query suggestions.
This difference in their traits is statistically significantly greater than expected by chance ($p<0.001$ with a two-tailed randomization test\footnote{We randomly swapped the system-generated gender labels among systems. Then for each system, we calculated the similarity score between the item list for a gender and the pre-documented item list for that gender and computed the difference between the similarity score of two systems}). This implies that these systems have the tendency of replicating gender stereotypes through query suggestions.

\textbf{Key Findings}: 
We have explored two e-commerce systems to identify if and how gender stereotypes manifest through query suggestions. Moreover, how the prevalence of gender stereotypes vary across these systems.

From our analysis of \textit{RQ1}, we have the following findings:
\begin{itemize}
    \item E-commerce search systems frequently target gender for children's items through query suggestions.
    \item The overall response (query suggestions) of all systems imply that gender stereotypes associated with children's products persist in search systems in e-commerce settings to at least some extent.
    \item The tendency of associating target gender with children's items vary across the selected systems.
    \item The tendency of reflecting previously-documented gender stereotypes associated with children's products through query suggestions vary between Amazon and Target.
    
\end{itemize}


\begin{table*}[]
\begin{minipage}{\textwidth}
\footnotesize
\centering
\caption{Top-10 most frequent unique words extracted from gender-specific search results}
\label{tab:freq_words}
\begin{tabular}{p{0.45in}>{\raggedright\arraybackslash}p{1.45in}>{\raggedright\arraybackslash}p{1.45in}>{\raggedright\arraybackslash}p{1.45in}>{\raggedright\arraybackslash}p{1.45in}}
\hline
\multicolumn{1}{l}{}                                    & \multicolumn{2}{c}{\textbf{Boys Only}}
& \multicolumn{2}{c}{\textbf{Girls Only}}                \\ \hline
\multicolumn{1}{l}{}                                    & \multicolumn{1}{c}{\textbf{Amazon}}                   & \multicolumn{1}{c}{\textbf{Target}}                   & \multicolumn{1}{c}{\textbf{Amazon}}                   & \multicolumn{1}{c}{\textbf{Target}}                   \\
\hline
\toprule
\textbf{Toys} 
& truck, car, dinosaur, construction, carrier, target, race, vehicles,  friction, arrow
& blaster, truck, nerf, vehicle, light, starter, kit, ipad, dart
& princess, makeup, tent, art, doll, unicorn, doodle, house, little, pink 
& surprise, baby, fashion, activity, unicorn, pet, puppy, dolls, hair, doll
\\
\midrule
\textbf{Games}                                                       & target, blaster, gun, dart, electronic, nerf, ball, blue, astroshot, magnetic 
& connect, boy, blue, foosketball, bulls, eye, sparkle, bash, baby, predictions 
& unicorn, dance, rainbow, crafts, women, music, bubble, candy, disney, princess 
& princess, meme, disney,  minnie, story, perfection, upwords, guess, magic, 
\\
\midrule
\textbf{Sports}
& toss, target, tee, electronic, challenge, boy magnetic, boxing, gloves, rocket  
& boys, nerf, red, football, playball, athletic, bat, ball, archery, game, 
& bra, big, women, beach, cotton, gym, yoga, tennis, crop, headband
& girls, bra, motion, purple, magic, slip, dye, cutout, cloud, strappy, 
\\
\midrule
\textbf{Books}
& awesome, big, boys, house, goodnight, tree, game, growth, national, knight 
& boy, big, hardy, bad, game, captain, underpants, battle, bionic, booger, 
& girl, unicorn, disney, creative, movement, story, young, confident, beautiful, mystery
& girl, stories, minute, power, brave, target, baby, frozen, story, disney
\\
\midrule
\textbf{Learning Material}
& card, gamenote, classroom, leapfrog, friends, frustration, green, control, scholastic, teaching 
& leapfrog, vtech, bilingual, words, book, real, sex, boys, gender, black
& pieces, homeschool, visual, tactile, auditory, mosaic, magnetic, fishing, wonderful, pattern
& activity, disney, princess, beads, kit, crinkle, paper, cloud, island, toy
\\
\midrule
\textbf{All Items}
& dinosaur, red, fire, blue, car, ball, constructions, vehicle, truck, race
& boy, war, toy, car, black, blue,  man, action, star, mega
& princess, pink, unicorn, purple, doll, disney, house, perfect, rainbow, white
& princess, girl, disney, frozen, mermaid, minnie, surprise, pink, fashion, doll
\\
\hline
\end{tabular}
\end{minipage}
\end{table*}

\subsection{Search Results}

After query suggestions, we wanted to explore the role of \emph{search results} in manifesting gender stereotypes in children's products. With that goal, we answered two research questions.  


Like before, we used items from Table~\ref{tab:items} to formulate queries, with the addition of  expanding each query to explicitly mention gender. For example, for ``action figures'', we searched for ``action figures for boys'' and ``action figures for girls'' in addition to the gender-neutral ``action figures for kids''.  We analyzed every search result for each item and used the title of retrieved results for content analysis.
\begin{figure}[tbh]
    \begin{subfigure}{\columnwidth}
    \centering
            \includegraphics[width=0.4\linewidth]{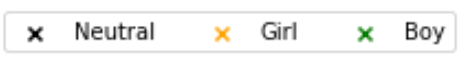}
    \end{subfigure}
    \hspace{1ex}
    \begin{subfigure}[t]{0.48\columnwidth}
    \includegraphics[width=\textwidth]{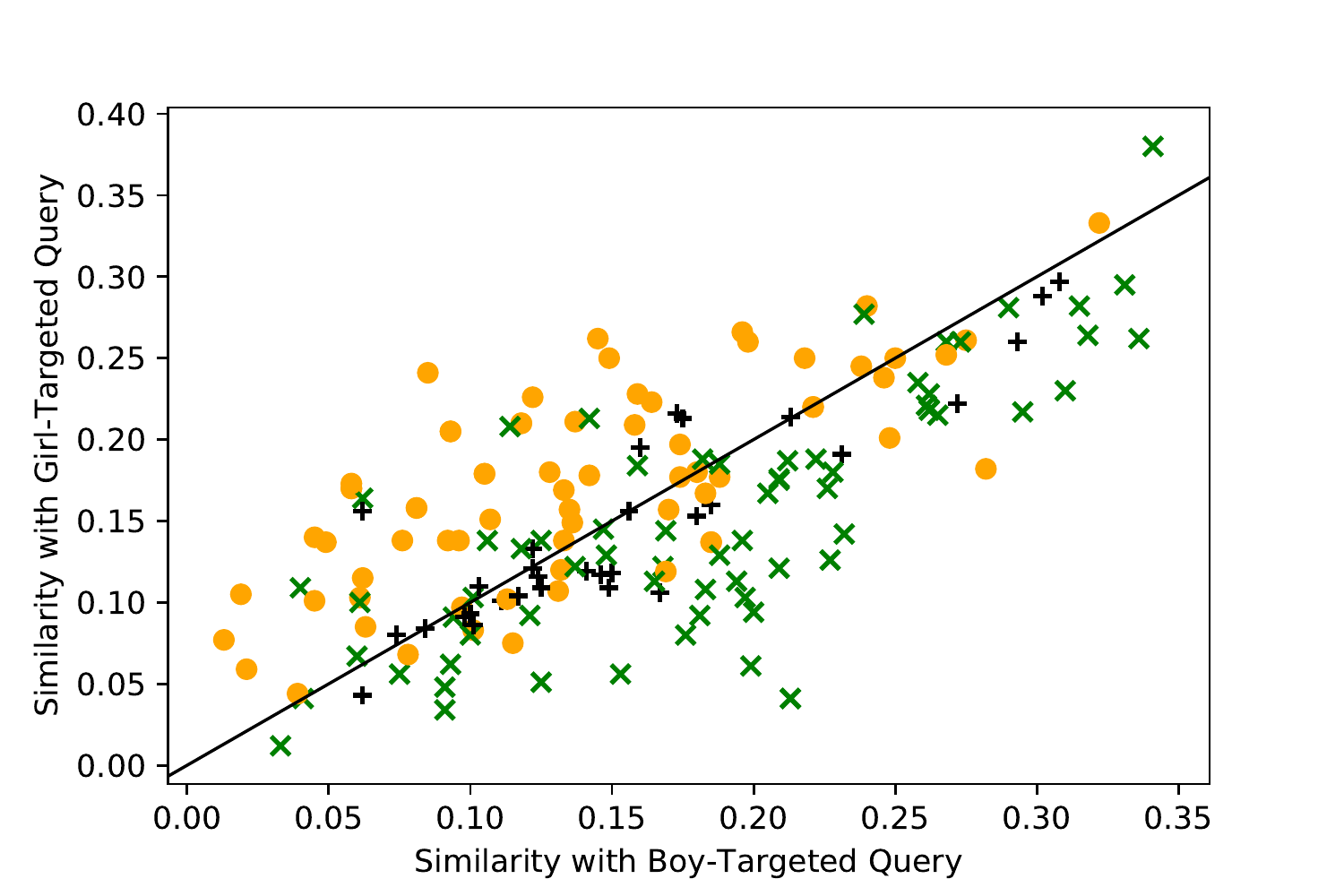}
    \caption{Amazon search results}
    \label{fig:AZ_scatter}
    \end{subfigure}
    \hspace{1ex}
    \begin{subfigure}[t]{0.48\columnwidth}
    \includegraphics[width=\textwidth]{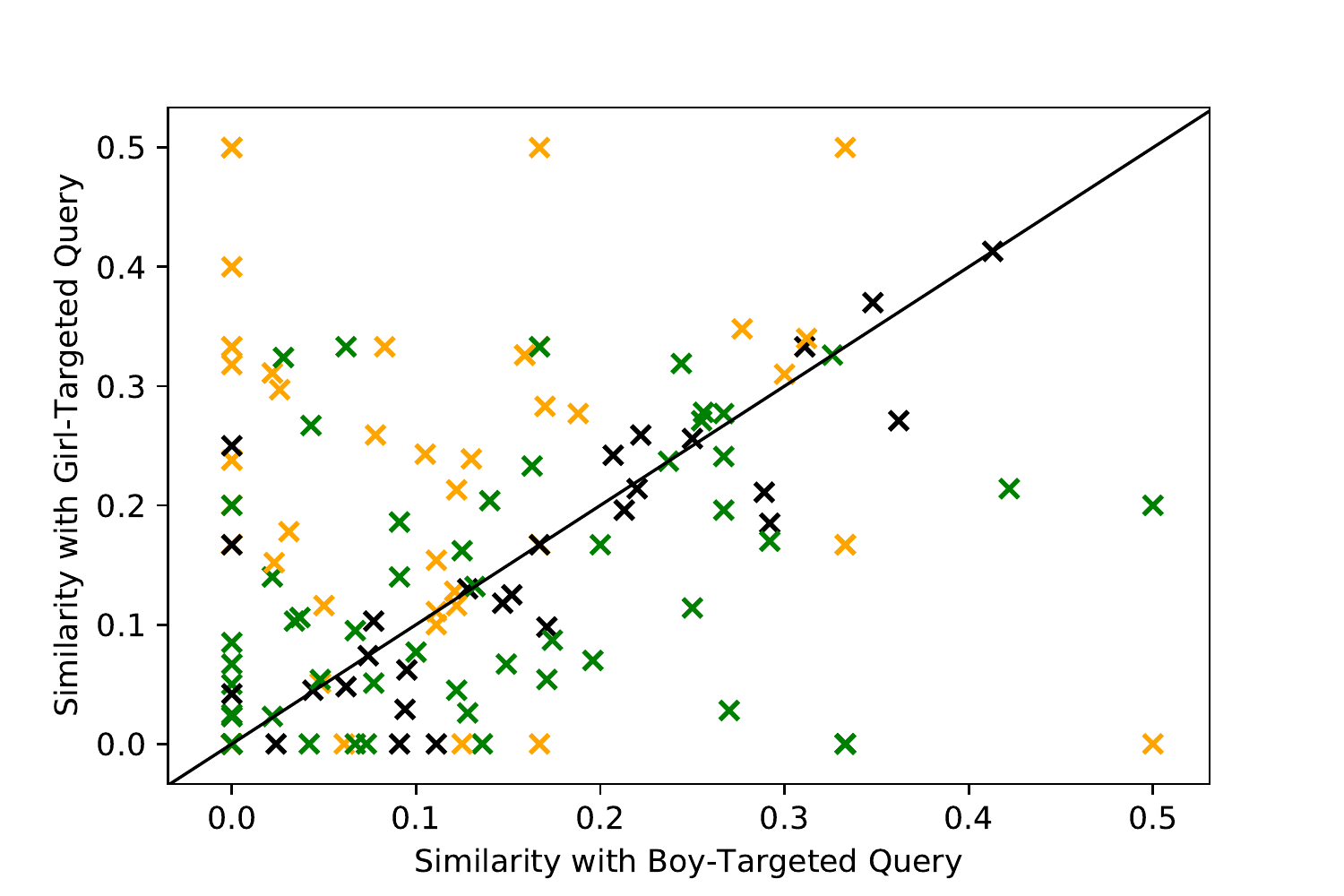}
    \caption{Target search results}
    \label{fig:target_scatter}
    \end{subfigure}
\caption{Gender associations for items in search results, colored by previously-documented stereotypes in our toy source lists. Yellow represents previously-documented girls' items, green are the previously-documented boys' items, and rest are neutral items.}
    \label{fig:gender_asso}
\end{figure}

\textit{\textbf{RQ2a.} How do search results show stereotypical responses to the gender mentioned in the query?}

For this analysis, we issued searches with our list of toys as well as generic searches (toys, books, sports, games, and learning materials). 
To detect if there is any difference in retrieved results with the change of gender in the query for the same item or generic search, we examined the terms frequently appearing in search results for only the gender specific queries.

We collected the titles of retrieved products and used \textit{NLTK} \citep{loper2002nltk} to filter the text, tokenized it (RegexpTokenizer), removed stopwords (NLTK's English corpus), and lemmatized the resulting terms (WordNetLemmatizer). We then identified nouns and adjectives using \texttt{pos\_tag} from \textit{NLTK} from the processed words. We counted the number of occurrence of each token in search results and identify unique keywords that only appeared with specific gendered search for an item (e.g. in ``toys for girls'' but not in ``toys for boys''). Table~\ref{tab:freq_words} shows the list of most-frequent terms associated with individual gendered search for each general query and for the set of all item queries, ordered by decreasing frequency. 

Previous studies on gender studies associated with children's products identified common traits in gender stereotyped products
\citep{murnen2016boys, cherney2018characteristics, coyne2016pretty, azmi2021gender}. For example, boy-targeted items often represent strong, action, assertiveness, violence, dominance, and aggressiveness. One the other hand, girl-targeted items often demonstrated domestic characteristics, soft or kind nature, and appearance focused \cite{blakemore2005characteristics}. Previous studies already established that "pink" is targeted for girls and "blue" for boys \cite{cherney2018characteristics}.
We have identified the persistence of the trend of associating stereotypes with gender in search results.

\begin{itemize}
    \item In both systems, ``toys for boys'' generates words that are typically represented only as masculine products i.e. ``truck'', ``car'', ``airplane'', ``arrow'', etc. and ``toys for girls'' generates feminine products like, ``princess'', ``makeup'', ``doll'', etc. The keywords from retrieved results show clear distinctions between genders which reflect the gender-stereotyped categorization of products in previous studies.
    \item In ``games for boys'', top results included violent terms like guns, bullet, blaster, battle, whereas frequent keywords like ``princess'', ``rainbow'', ``unicorn", associated with ``games for girls'' show attractive characteristics \cite{fawcett19, coyne2016pretty}.
    \item We can see further distinctions in ``sports''. ``sports for boys'' elicits keywords that are related to different sports and activities, whereas ``sports for girls'' mostly represents appearance-based keywords like clothing accessories, consistent with the findings of \citet{murnen2016boys} that boys' items are generally action-based whereas girls' items are more appearance-based.
    \item Gendered search for ``books'' shows the similar trend by using adventurous terms to represent ``books for boys'' and magical terms to represent ``books for girls'' \cite{lewis2020might}.
    \item When searching for specific items instead of broad categories, we still saw gender-specific result keywords. ``pink'' is the most common terms for representing items for girls and ``blue'' for boys. Disney princesses like Elsa, Ariel, Belle appeared frequently in girl-targeted results whereas superheroes like Batman and Superman frequently appeared in boy-targeted results.
    \item In addition, there is a difference in adjectives in product representation while targeting certain genders. ``adventure'', ``action'', ``electronic'', ``magnetic'', ``strategic'', and ``crazy'' are frequently found in boy-targeted products, whereas, ``beautiful'', ``magical'', ``mystery'', ``musical'', and ``colorful'' are most commonly used only for girls.
\end{itemize}

These differences in gender-targeted search results qualitatively indicates that the selected e-commerce search systems replicated and reinforced gender stereotypes through search results in order to target a specific gender, in ways that sometimes correspond to specific previously-documented stereotype associations.

\textit{\textbf{RQ2.b} Do e-commerce search systems associate items with genders through search results?}

This analysis uses all three of the queries for each toy (gender-neutral plus the two gendered queries) for each item in Table~\ref{tab:items}. 
We then investigated the differences in sets of items retrieved, instead of their keywords, as we changed the gender of the query, which gender the search engine associates more strongly with that toy category. We do this by measuring the similarity between the gender-neutral result set and each of the gender-targeted result sets; if the neutral search results closely match with a specific gendered search results, we identified that product to be associated with that gender. For example, if search results for \textit{``dolls for kids''} has more items in common with search results for \textit{``dolls for girls''} than \textit{``dolls for boys''}, we take that as evidence that the system associates dolls more strongly with girls than boys.

We measured the Jaccard similarity between the set of items retrieved by the gender-neutral query and the sets produced by boy-specific and girl-specific queries, resulting in Boy and Girl scores for each query on each systems. 
\begin{table*}[tbh]
\caption{Top-20 gender specific items assumed by observed systems along with their similarity score with gender-neutral search results. Items are sorted based on difference between similarity Score of boy-Specific and girl-Specific results. }
\label{tab:similarity}
\footnotesize
\begin{tabular}{llllllllll}
\hline
\multicolumn{5}{c}{\textbf{Amazon}}                                                                                                                                                                                          & \multicolumn{5}{c}{\textbf{Target}}                                                                                                                                                                                          \\ \hline
\multicolumn{1}{c}{\textbf{Item}} & \multicolumn{2}{c}{\textbf{Similarity Score}}                        & \textbf{Difference} & \multicolumn{1}{c}{\textbf{Prev. Label}} & \multicolumn{1}{c}{\textbf{Item}} & \multicolumn{2}{c}{\textbf{Similarity Score}}                        & \textbf{Difference} & \multicolumn{1}{c}{\textbf{Prev. Label}} \\ \hline
\multicolumn{10}{c}{\textbf{Top 20 Boy-Aligned Items}}                                                                                                                                                                                                                                                                                                                                                                                                               \\
                                  & \multicolumn{1}{c}{\textbf{Boy}} & \multicolumn{1}{c}{\textbf{Girl}} & \textbf{Boy-Girl}   & \textbf{}                                                                                   & \multicolumn{1}{c}{\textbf{}}     & \multicolumn{1}{c}{\textbf{Boy}} & \multicolumn{1}{c}{\textbf{Girl}} & \textbf{Boy-Girl}   &                                                                                             \\
military toys                     & 0.21                             & 0.04                              & 0.17                & Boy                                                                                         & power ranger valentine            & 0.5                              & 0.0                               & 0.50                & Girl                                                                                        \\
football                          & 0.2                              & 0.061                             & 0.13                & Boy                                                                                         & toy drones                        & 0.33                             & 0.0                               & 0.33                & Boy                                                                                         \\
outer-space toys                  & 0.28                             & 0.15                              & 0.13                & Boy                                                                                         & toy robots                        & 0.33                             & 0.0                               & 0.33                & Boy                                                                                         \\
doll humanoid                     & 0.2                              & 0.09                              & 0.10                & Boy                                                                                         & monster trucks                    & 0.33                             & 0.0                               & 0.33                & Boy                                                                                         \\
hockey stick                      & 0.15                             & 0.05                              & 0.10                & Boy                                                                                         & nerf guns                         & 0.33                             & 0.0                               & 0.33                & Boy                                                                                         \\
spiderman                         & 0.15                             & 0.05                              & 0.10                & Boy                                                                                         & soldier toys                      & 0.31                             & 0.0                               & 0.31                & Boy                                                                                         \\
wwe action figure                 & 0.176                            & 0.08                              & 0.09                & Boy                                                                                         & dinosaur toys                     & 0.5                              & 0.2                               & 0.30                & Boy                                                                                         \\
superhero costume                 & 0.17                             & 0.07                              & 0.09                & Boy                                                                                         & ninja costume                     & 0.27                             & 0.02                              & 0.24                & Boy                                                                                         \\
spiderman costume                 & 0.19                             & 0.10                              & 0.09                & Boy                                                                                         & pokemon cards                     & 0.42                             & 0.24                              & 0.20                & Boy                                                                                         \\
sword toy                         & 0.23                             & 0.14                              & 0.09                & Boy                                                                                         & baking kit                        & 0.16                             & 0.0                               & 0.16                & Girl                                                                                        \\
construction toys                 & 0.18                             & 0.09                              & 0.09                & Boy                                                                                         & princess sword                    & 0.33                             & 0.16                              & 0.16                & Girl                                                                                        \\
dinosaur toy                      & 0.20                             & 0.12                              & 0.08                & Boy                                                                                         & airport toys                      & 0.13                             & 0.0                               & 0.16                & Boy                                                                                         \\
batman costume                    & 0.19                             & 0.11                              & 0.08                & Boy                                                                                         & construction toys                 & 0.19                             & 0.07                              & 0.16                & Boy                                                                                         \\
space station toys                & 0.31                             & 0.23                              & 0.08                & Boy                                                                                         & cupcake maker                     & 0.13                             & 0.0                               & 0.13                & Girl                                                                                        \\
weather forecasting toys          & 0.29                             & 0.21                              & 0.08                & Boy                                                                                         & remote controlled car             & 0.3                              & 0.17                              & 0.13                & Boy                                                                                         \\
miniature guns                    & 0.33                             & 0.26                              & 0.07                & Boy                                                                                         & gi joe action figure              & 0.17                             & 0.05                              & 0.12                & Boy                                                                                         \\
action figures                    & 0.13                             & 0.05                              & 0.07                & Boy                                                                                         & winnie the pooh                   & 0.11                             & 0.0                               & 0.11                & Neutral                                                                                     \\
toy animals                       & 0.16                             & 0.1                               & 0.07                & Boy                                                                                         & playdoh                           & 0.29                             & 0.18                              & 0.10                & Neutral                                                                                     \\
ninja costume                     & 0.19                             & 0.12                              & 0.07                & Boy                                                                                         & racetrack                         & 0.12                             & 0.02                              & 0.10                & Boy                                                                                         \\
grill                             & 0.19                             & 0.13                              & 0.06                & Boy                                                                                         & wagon                             & 0.09                             & 0.0                               & 0.09                & Boy                                                                                         \\
\multicolumn{10}{c}{\textbf{Top 20 Girl-Aligned Items}}                                                                                                                                                                                                                                                                                                                                                                                                              \\
vanity set                        & 0.08                             & 0.24                              & -0.15               & Girl                                                                                        & fashion doll                      & 0.0                              & 0.5                               & -0.50               & Girl                                                                                        \\
play makeup                       & 0.14                             & 0.26                              & -0.11               & Girl                                                                                        & disney doll                       & 0.0                              & 0.5                               & -0.50               & Girl                                                                                        \\
pink doll                         & 0.05                             & 0.17                              & -0.11               & Girl                                                                                        & toy helicopter                    & 0.02                             & 0.41                              & -0.40               & Boy                                                                                         \\
beads                             & 0.05                             & 0.17                              & -0.11               & Girl                                                                                        & tiara                             & 0.0                              & 0.4                               & -0.40               & Girl                                                                                        \\
dollhouse                         & 0.09                             & 0.2                               & -0.11               & Girl                                                                                        & domestic toys                     & 0.0                              & 0.33                              & -0.33               & Girl                                                                                        \\
dorothy costume                   & 0.12                             & 0.23                              & -0.10               & Girl                                                                                        & tea party set                     & 0.16                             & 0.5                               & -0.33               & Girl                                                                                        \\
car toys                          & 0.06                             & 0.16                              & -0.10               & Boy                                                                                         & princess costume                  & 0.0                              & 0.32                              & -0.31               & Girl                                                                                        \\
makeup kit                        & 0.14                             & 0.25                              & -0.10               & Girl                                                                                        & sully costume                     & 0.02                             & 0.32                              & -0.29               & Girl                                                                                        \\
unicorn                           & 0.04                             & 0.14                              & -0.09               & Girl                                                                                        & beads                             & 0.2                              & 0.31                              & -0.28               & Girl                                                                                        \\
drum set                          & 0.06                             & 0.15                              & -0.09               & Neutral                                                                                     & remote controlled helicopter      & 0.06                             & 0.33                              & -0.27               & Boy                                                                                         \\
castle tent                       & 0.11                             & 0.2                               & -0.09               & Girl                                                                                        & barbie costume                    & 0.02                             & 0.29                              & -0.27               & Girl                                                                                        \\
baby doll stroller                & 0.11                             & 0.21                              & -0.092              & Girl                                                                                        & spongebob squarepants             & 0.0                              & 0.25                              & -0.25               & Neutral                                                                                     \\
painting kit                      & 0.05                             & 0.14                              & -0.08               & Girl                                                                                        & ken doll                          & 0.08                             & 0.33                              & -0.25               & Girl                                                                                        \\
jewelry                           & 0.01                             & 0.1                               & -0.08               & Girl                                                                                        & my little pony                    & 0.0                              & 0.23                              & -0.23               & Girl                                                                                        \\
baby doll                         & 0.08                             & 0.15                              & -0.07               & Girl                                                                                        & wwe action figure                 & 0.043                            & 0.26                              & -0.22               & Boy                                                                                         \\
my little pony                    & 0.1                              & 0.21                              & -0.07               & Girl                                                                                        & lego cars                         & 0.0                              & 0.2                               & -0.20               & Boy                                                                                         \\
sully costume                     & 0.14                             & 0.21                              & -0.07               & Girl                                                                                        & craft toys                        & 0.0                              & 0.2                               & -0.20               & Girl                                                                                        \\
tea party set                     & 0.19                             & 0.27                              & -0.07               & Girl                                                                                        & vanity set                        & 0.07                             & 0.26                              & -0.18               & Girl                                                                                        \\
tea set                           & 0.15                             & 0.22                              & -0.07               & Girl                                                                                        & baby doll stroller                & 0.15                             & 0.32                              & -0.16               & Girl                                                                                        \\
doll                              & 0.01                             & 0.07                              & -0.069              & Girl                                                                                        & toy animals                & 0.0                             & 0.16                              & -0.16               & Girl                                                                                        \\ \hline
\end{tabular}
\end{table*}
Table~\ref{tab:similarity} presents the top 20 individual boy- and girl-aligned items, based on the difference in boy and girl similarity scores. We can see that item lists vary between systems.
Figure~\ref{fig:AZ_scatter} and ~\ref{fig:target_scatter} show the derived gender affinity match with the previously-documented gendered item list. Both systems displayed a tendency of identifying items as gender-specific reflecting the previously-documented gender stereotypes to a significant extent: 97\% of the top gender-aligned items from Amazon match with the previously-documented gender label of these items, while 72\% of the top Target items do. 
This indicates the potential of reinforcing and preserving gender stereotypes in children's products through search results.

\textbf{Key findings:} In \textbf{RQ2}, we explored the existence of gender stereotypes in search results for children's products. We made the following observations:
\begin{itemize}
    \item For the same item, retrieved results change with the presence of gender in the query.
    \item Products are often associated with gender stereotyped keywords. IR systems reflect and reinforce that pattern by manifesting stereotypical gender-typed contents through retrieved results targeting a specific gender.
    \item Both systems show a similar tendency in replicating and displaying gender stereotypes through search results.
\end{itemize}

\section{Discussion}

Through analyzing query suggestions, we observed that production e-commerce systems often associates gender with items (by suggesting gender-related queries when searching for the item), and that these associations reflect the common social practice of categorizing children's products by gender. Although the degree of this tendency of gender association varies across different systems, no e-commerce system in our experiment showed complete exception from this behavior. By suggesting target gender with children's items, these systems reproduce previously-documented gender stereotypes users' search experiences and may influence them to select accordingly.

Content analysis of search results allowed us to explore if and how e-commerce search systems manifest gender stereotypes in children's items through search results.  
We observed that content representation changes with presence of gender in the query, supporting pre-defined gender-specific item characteristics.
This finding implies that e-commerce search systems had the tendency of categorizing items based on gender and promoting stereotypical gendered target of children's products. 

The similarity between gender specific search and gender neutral search for same item let us identify if e-commerce search systems have the potential to manifest gender stereotypes by assuming item appropriateness for a specific gender. Our results from analyzing two of the most popular e-commerce sites show that e-commerce search systems can infer gender affiliation with items and retrieve relevant information accordingly which indicates the possibility of these systems reinforcing the practice of gender-targeted children's products marketing.

Our findings provide evidence that e-commerce search engines can reproduce gender stereotypes through query suggestions and search results, and we present initial methods for quantifying this effect. This demonstrates the importance of studying e-commerce platforms, both in their explicit design and emerging properties of their supporting systems.
Even when explicit gender categorizations are removed from physical stores and e-commerce site layouts, as Target has done, a site's search engine may still carry latent categorization that can lead the system to replicate gender stereotypes. 

\section{Limitations and Future Directions}
In this paper, we have provided initial evidence that search systems can propagate stereotypical gender associations into search results, particularly for children's products.

As an early exploratory study, there are a number of limitations we acknowledge in this work. There is much more work to do to fully and more precisely understand this phenomenon, and we hope our study provides a useful foundation and motivation for future work that addresses many of these limitations.

Our toy sets lists are based on lists compiled by previous research studies and advocacy campaigns to identify toys with gender stereotyped associations; this collection is not necessarily representative and is limited in size. New data collection strategies, possibly based on surveys, to identify a wider range of products and their stereotypical gender associations will enable a more thorough and representative study of the phenomenon.

We then directly used the items as search queries, because logs of actual queries issued by users of an e-commerce system are often not publicly available, particularly when children might be searching. Collecting and analyzing actual queries, particularly queries written by children, would help us better measure system responses to their actual use. User studies with children, parents, and educators would provide understanding of their perception and preference about gendered-typed toys and the concerns of early childhood exposure of gender stereotypes through IR systems. 

Our content analysis focuses on retrieved product titles and extracted keywords; more in-depth analysis that uses more content (product images, descriptions, etc.) and more sophisticated analysis techniques are welcome.
Also, by directly measuring the similarity in retrieved products between gender-specific and gender-neutral, we identified system tendency to associate items with gender through search results. This has enabled us to gather evidence of the extent of stereotype reproduction, but more work is needed to refine and validate measurement methodologies. In the future, we can work towards more fine grained and conclusive strategies to identify gender stereotypes in search results.

Our main focus for this work was on e-commerce search systems considering two vital components of these systems: query suggestions and search results for our analysis. However, the role of recommendations in propagating gender stereotypes associated with children's products remains open to explore in the future.

Finally, our search result findings are limited to two popular e-commerce sites; other systems may well represent stereotypes in different ways (or not at all).

Additional directions of future research regarding this issue include more refined methods to measure gender stereotypes in IR systems, identifying and understating the causes of these stereotypes, studying their impact, and developing mitigation strategies.


Many researchers and advocates are concerned about gender stereotypes in children's early experiences, particularly through the toys they play with. We hope this work will prove a useful foundation for researchers to join us in understanding the role IR systems play and how they may affect children's development.


\begin{acks}
This material is based upon work supported by the National Science Foundation under Grant No. IIS 17-51278. We also thank our colleagues in the People and Information Research Team (PIReT) for their ongoing support, and particularly Dr. Sole Pera for her valuable feedback and guidance on this project.
\end{acks}

\bibliographystyle{ACM-Reference-Format}
\bibliography{reference}


\end{document}